\documentclass[12pt]{amsart}
\usepackage{geometry}                
\usepackage{graphicx}
\usepackage{amssymb}
\usepackage{epstopdf}
\title{Deterministic chaos in government debt dynamics with mechanistic primary balance rules}
\author{Jussi Ilmari Lindgren}

\begin{document}
\maketitle

\begin{abstract}
This paper shows that with mechanistic primary budget rules and with some simple assumptions on interest rates the well-known debt dynamics equation transforms into the infamous logistic map. The logistic map has very peculiar and rich nonlinear behaviour and it can exhibit deterministic chaos with certain parameter regimes. Deterministic chaos means the existence of the butterfly effect which in turn is qualitatively very important, as it shows that even deterministic budget rules produce unpredictable behaviour of the debt-to-GDP ratio, as chaotic systems are extremely sensitive to initial conditions. 
\end{abstract}
\section{Introduction}
Deterministic chaos and economics do not have very fruitful relationship in terms of scientific synergy. Nevertheless, it is important to acknowledge that the possibility of chaos in economic phenomena has some important consequences. The essence of chaotic dynamical systems is that they are essentially unpredictable. Minor changes in the initial conditions will lead to drastic changes in the system dynamics. In Keynesian economic policy it is usually assumed that the government can control the economy by providing additional demand during bad times, which would ideally dampen at least up to some extent the unfavorable business cycles. This study basically shows that even with very simplistic and deterministic fiscal policy rules the debt dynamics will become chaotic and therefore unpredictable. The initial conditions of the system are always uncertain in the real world, which guarantees that the dynamics are therefore subject to the so-called butterfly effect. There has been relatively little research on chaos and economics. There are few exceptions, however. Most notably Matti Pohjola \cite{mp} has studied the chaotic dynamics of the famous Goodwin model of class struggle\cite{goodwin}. The nonlinear model presented here also provides insight what kind of fiscal policy rules can stabilize the debt-to-GDP ratio without any additional assumptions on the economy as a whole.  The model presented here is not based on the methodology of the current mainstream approach of Dynamic Stochastic General Equilibrium, but instead on the simple dynamics of the debt which depends on the interest rates, economic growth and the primary balance. It is especially interesting to notice that small changes in these parameters will change the behavior of the system drastically. 
\section{The Model}
The well-known debt dynamics in per capita form is given by the difference equation
\begin{equation}
b_{t+1}=p_{t}+\frac{1+i}{1+g}b_{t}
\end{equation}
The government debt-to-GDP ratio is given by $b_t$ and the rate of GDP growth is given by $g$.
In the equation $p_{t}$ denotes the primary deficit, if $p_{t}\geq 0$ and primary surplus, if $p_{t}\leq 0$. The interest rate on the debt is $i$.
Suppose now that the primary balance is governed by the mechanistic rule
\begin{equation}
p_{t}=\frac{\alpha b_{t}}{1+g}+\frac{\beta b_{t}^2}{1+g}
\end{equation}
where $\alpha$,$\beta$ are real parameters. 
Assume now that the interest rate on the debt is dependent on the debt-to-GDP ratio in the following way:
\begin{equation}
i=\gamma b_{t}
\end{equation}
with some $\gamma \geq 0$. In this study we shall assume $\gamma =0,1$. This is very reasonable assumption as for example country with 100 per cent indebtness would face an interest rate of 10 per cent on its debt. Similarly a country with 50 per cent indebtness would face an interest rate of 5 per cent. It is important to notice that the primary balance rule will depend on the sensitivity of the yield. It is assumed that the government in question has full information on this parameter and adjusts its reaction function accordingly.

This means that the sensitivity of the yield is
\begin{equation}
\frac{di}{db_{t}}=\gamma
\end{equation}
Substituting the budget rule and the interest rate in the debt dynamics equation, one has
\begin{equation}
b_{t+1}=\frac{\alpha b_{t}}{1+g}+\frac{\beta b_{t}^2}{1+g}+\frac{b_{t}}{1+g}+\frac{\gamma b_{t}^2}{1+g}
\end{equation}
After simplifying, one gets
\begin{equation}
b_{t+1}=\frac{b_t}{1+g}\left (\alpha +\beta b_t +1 +\gamma b_t \right )
\end{equation}
\begin{equation}
b_{t+1}=\frac{b_t}{1+g}\left (\alpha +1+(\beta + \gamma )b_t \right )
\end{equation}
\begin{equation}
b_{t+1}=\frac{\beta + \gamma}{1+g}b_{t}\left (\frac{\alpha +1}{\beta + \gamma}+b_t \right )
\end{equation}
Assume now that the parameters obey
\begin{equation}
\frac{\alpha +1}{\beta + \gamma}=-1
\end{equation}
This means that the parameters $\alpha$ and $\beta$ are dependent on the market sensitivity $\gamma$
\begin{equation}
\alpha + \beta =-\gamma -1
\end{equation}
Therefore it is assumed that the government observes the current market sensitivity of yields and this will restrict the range of the two parameters in the primary balance rule. 
Then one has
\begin{equation}
b_{t+1}=\frac{-(\beta +\gamma)}{1+g}b_{t}\left (1-b_t \right )
\end{equation}
This is the infamous logistic map that permits very complex behavior and is an example of a simple dynamical system that can exhibit deterministic chaos. A very good introduction to the logistic map and chaotic dynamics is given in \cite{kaaos}We shall make the following convention
\begin{equation}
r=\frac{-(\beta +\gamma)}{1+g}
\end{equation}
We shall assume all the time that $g=0$ and $\gamma =0.1$. One should note that the market sensitivity and rate of economic growth will have major consequences in the qualitative behavior of the model.
\section{Debt dynamics with $0\leq r \leq 1$}
Let us first assume that $r=1$. This must imply then $\beta =-1.1$ and $\alpha =0$. The fiscal policy rule then becomes
\begin{equation}
p_{t}=-1.1b_{t}^2
\end{equation}
The logistic map has the feature that with any initial debt ratio $b_0 \in [0,1]$with $r\leq 1$ will lead to the the limit $b_t \longrightarrow 0$ as $t \longrightarrow \infty$.
The following figure illustrates the dynamics of the government debt with $r=1$ and $b_{0}=0.3$. 
\includegraphics[scale=0.8]{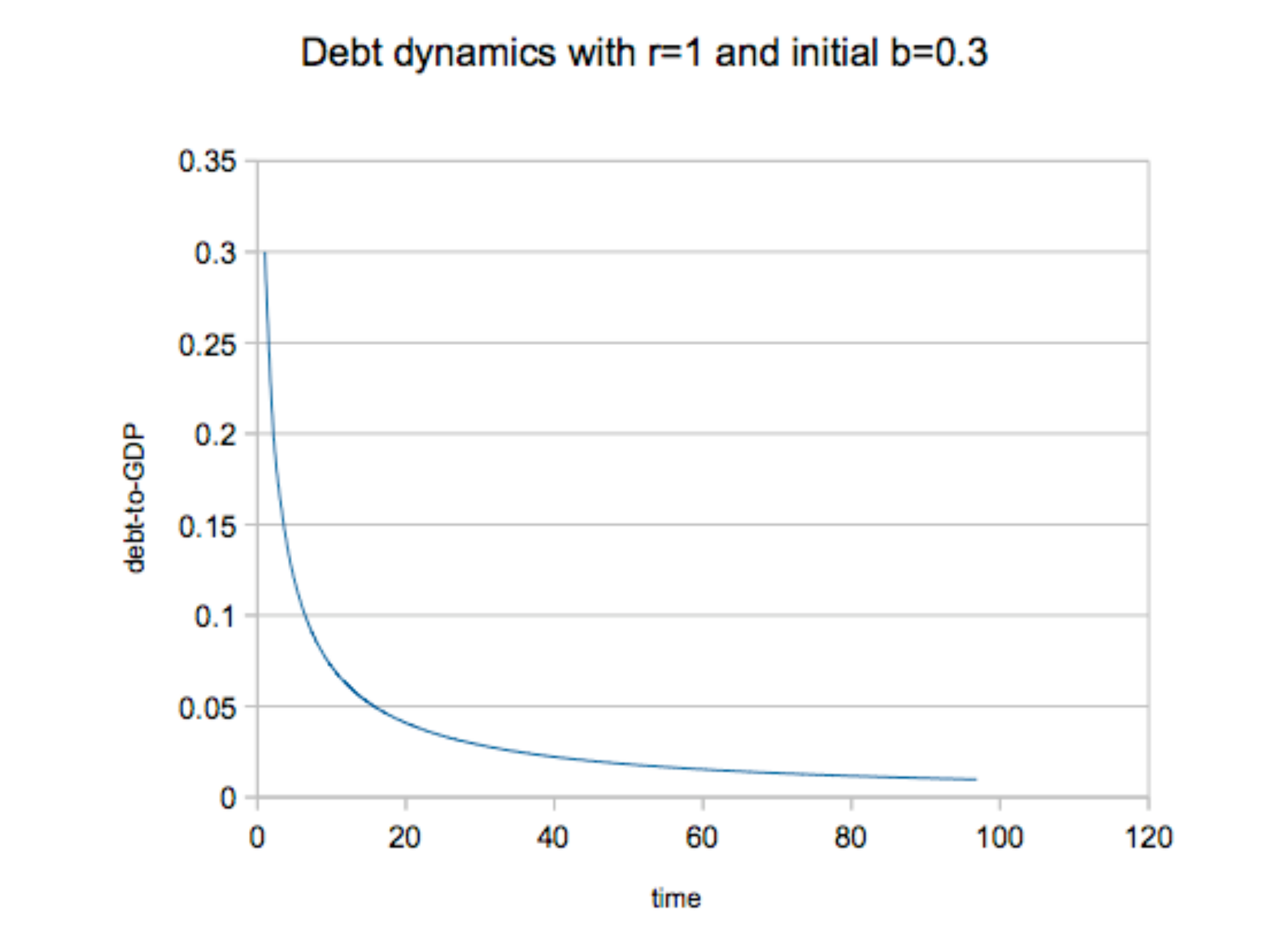}
\section{Debt dynamics with $r=3$}
Assume now that $r=3$. The logistic map has the feature that with $r\leq 3$ there is always a fixed point/s, in other words the debt-to-GDP ratio will eventually stabilize at some level or will oscillate around some region. The assumption implies that $\beta =-3.1$ and $\alpha =2$. This means that the fiscal policy rule is
\begin{equation}
p_{t}=2b_t-3.1b_{t}^2
\end{equation}
\includegraphics[scale=0.8]{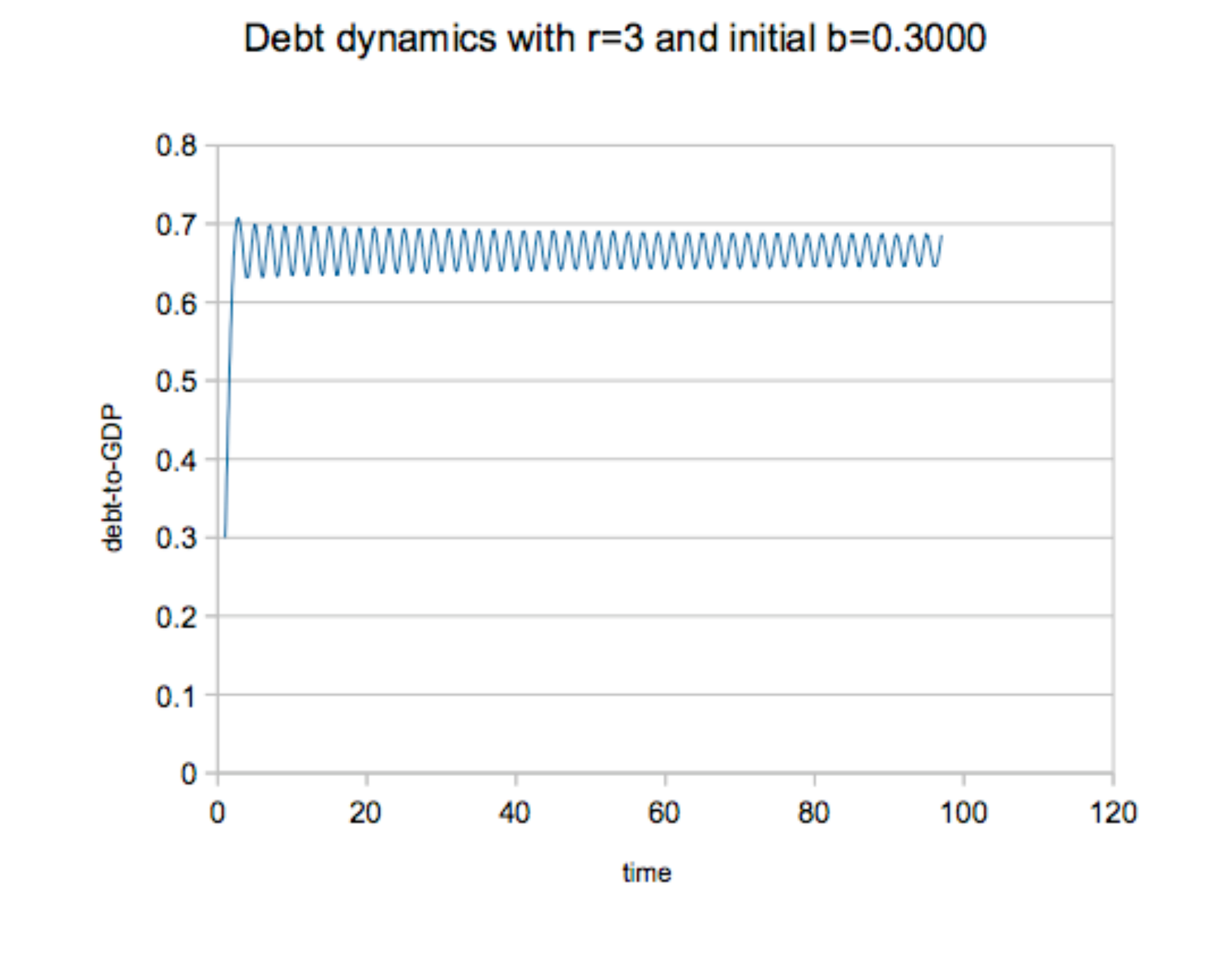}
\section{Debt dynamics with $r=3,5$}
Let us now assume that $r=3,5$. This implies with $g=0$ and $\gamma =0,1$ that $\beta=-3,6$ and $\alpha =2,5$. The fiscal policy rule is now
\begin{equation}
p_t=2,5b_t-3,6b_{t}^{2}
\end{equation}
\includegraphics[scale=0.9]{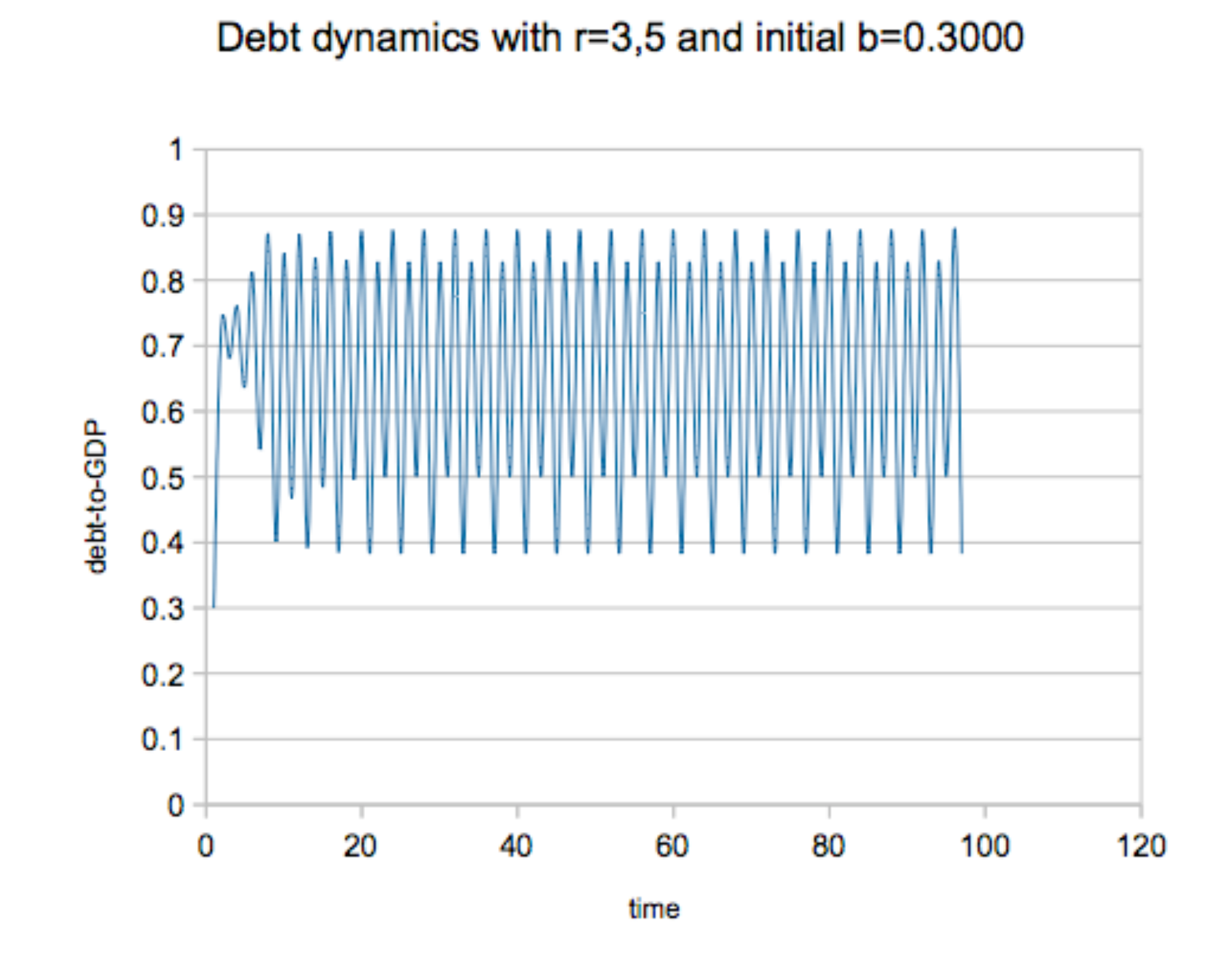}

Assume now that the economy is shrinking with the rate of $g=0,05$. Then $r\approx 3,684$ and the debt dynamics will be drastically different

\includegraphics[scale=0.9]{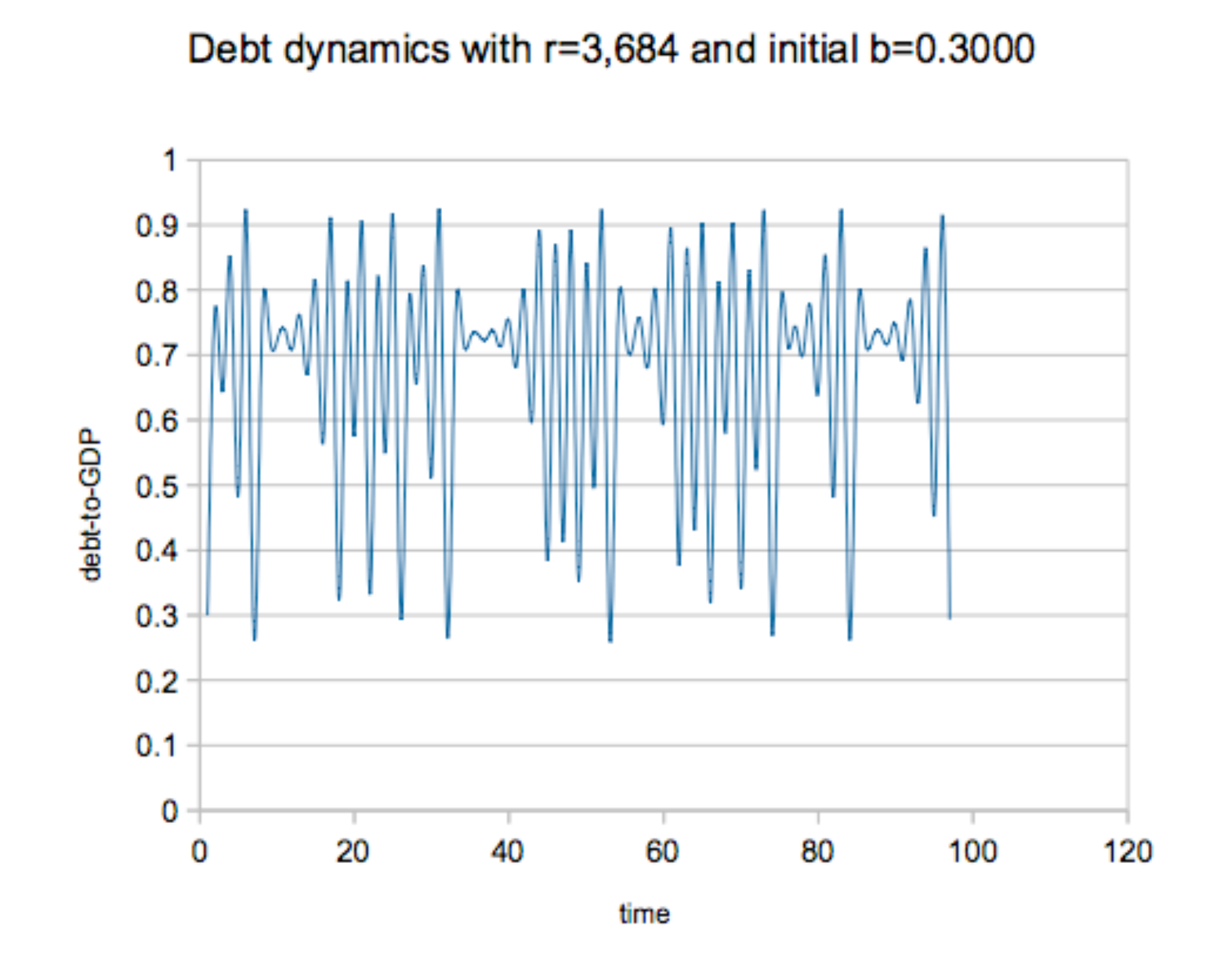}

The system dynamics is now chaotic and minor changes in the initial data will affect the dynamics in large proportions, consider the same dynamic with initial debt of $0,2$ instead of $0,3$.

\includegraphics[scale=0.9]{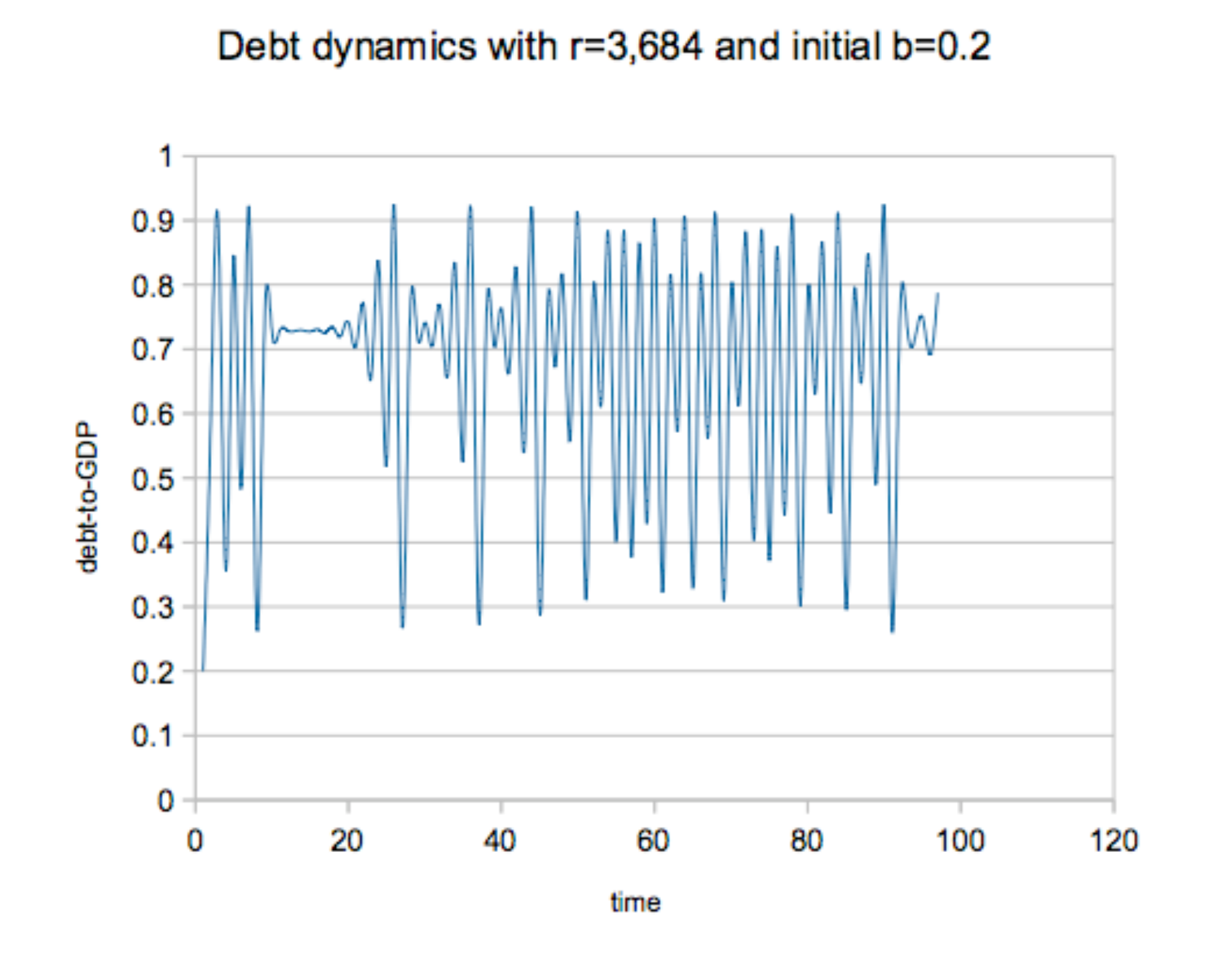}

The final debt level is now $b_{100}\approx 0,8$ whereas with initial debt of $0,3$ the final debt level is approximately $0,3$. The maximal primary surplus is approximately 40 per cent.
\section{Debt dynamics with $r=4$}
This upper bound for the parameter, $r=4$, implies full chaos in the debt dynamics. With $r$ greater than 3 the system becomes complicated. There are bifurcations and after a critical value of $r$ the system goes into the chaotic regime. This occurs approximately at $r=3,569946...$. The system is now extremely sensitive to initial conditions, hence exhibiting the 'butterfly effect'. For $r=4$ it implies that $\beta =-4.1$ and $\alpha =3$. The initial condition will vary in the following figures slightly around $0.3$. Notice the drastic structural differences when the initial debt differs only insignificantly.

\includegraphics[scale=0.8]{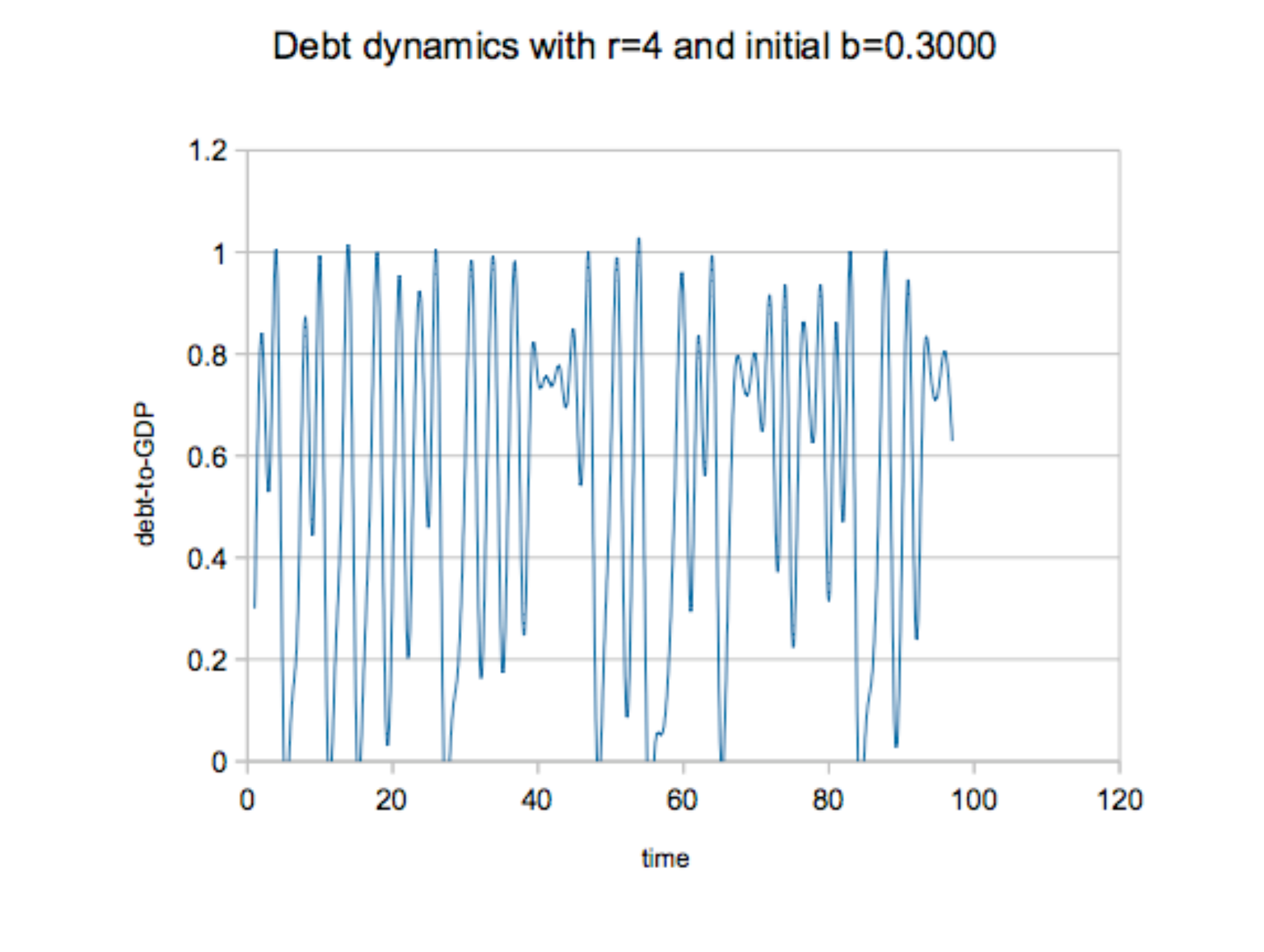}
\includegraphics[scale=0.8]{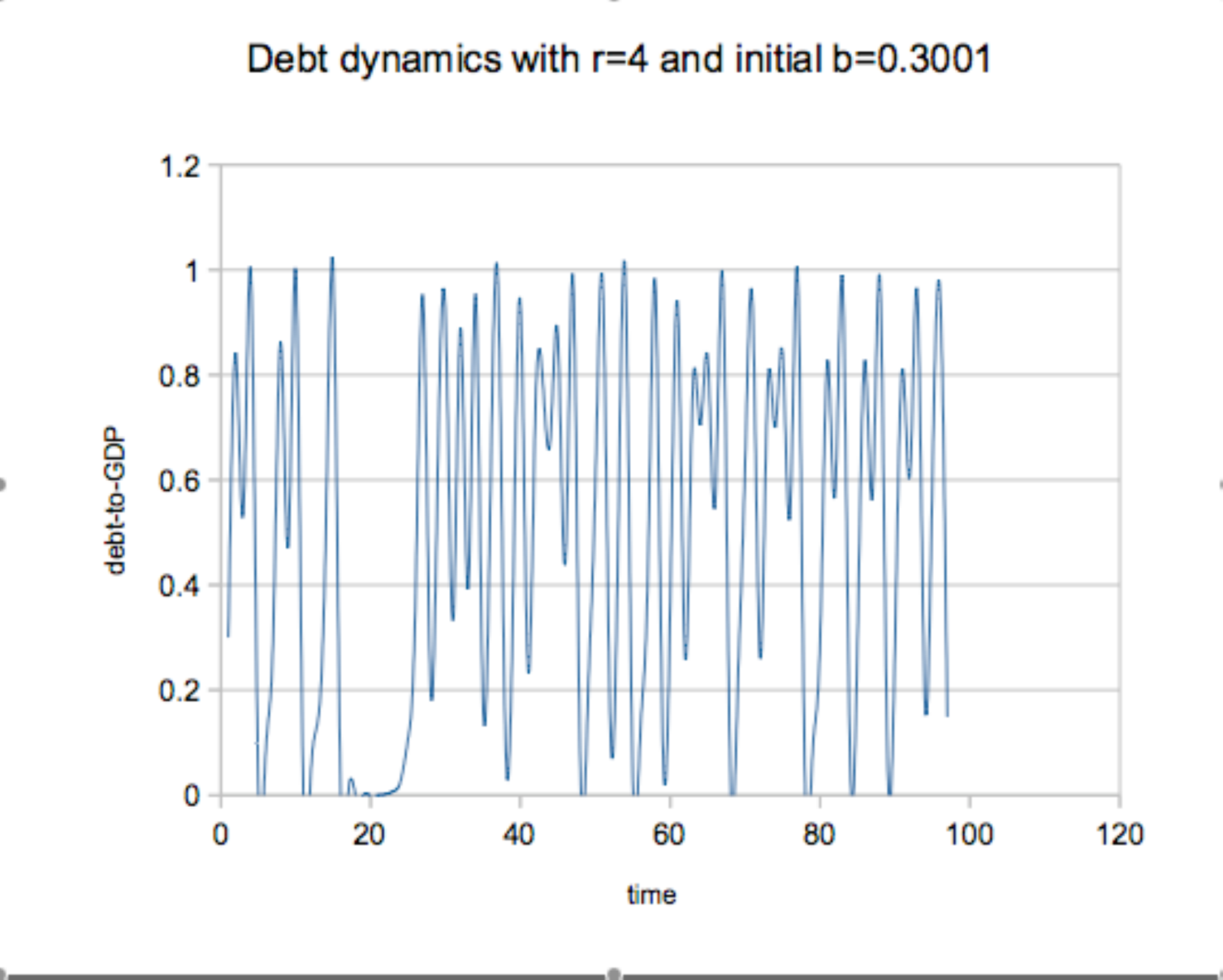}
Assume now for simplicity that $g=0$ and $\gamma =0,1$. Additionally we assume that $\alpha =3$ and $\beta=-4,1$.  This means that the primary balance obeys
\begin{equation}
p_t=3b_t-4,1b_{t}^2
\end{equation}
The local maximum occurs at $b=\frac{3}{8,2}=37$ per cent, which in turn means a maximum primary deficit of approximately 55 per cent.
Accordingly the debt dynamics equation reduces to
\begin{equation}
b_{t+1}=4b_{t}\left (1-b_{t}\right )
\end{equation}
Then the debt-to-GDP is bounded to the interval $[0,1]$ and the maximum primary surplus is 110 per cent when $b_{t}=1$.
\section{Conclusions}
The model presented here shows that for different rates of GDP-growth, market sensitivity and primary balance rules the debt dynamics differ vastly in structure. For some values of the parameters the system has a fixed point and for others the system is chaotic. This means that even with simple rules the governments' s debt position can be unpredictable. This in turn means that fiscal policy and debt stabilization is a nontrivial issue and the public finances cannot be controlled easily or even at all in the chaotic regime.

\end{document}